\documentstyle[twocolumn,prb,aps,psfig,epsf,fleqn,ifthen]{revtex}

\begin{document} 
 

\title{Observation of Andreev Reflection Enhanced Shot Noise} 
 
\author{P. Dieleman\cite{dank}, H.G. Bukkems, T.M. Klapwijk} 
\address{Department of Applied Physics and Materials Science Center, University of Groningen, Nijenborgh 4, 
9747 AG Groningen, The Netherlands}  
 
\author{M. Schicke, and K.H. Gundlach} 
\address{Institut de Radio Astronomie Millim\'{e}trique, 300 Rue de la Piscine, Domaine Universitaire de 
Grenoble,38406 St. Martin d'H\`{e}res, France.}

\maketitle 
 
\begin{abstract} 
We have experimentally investigated the quasiparticle shot noise in NbN/MgO/NbN superconductor - insulator -
superconductor tunnel junctions. The observed shot noise is significantly larger than theoretically expected. We 
attribute this to the occurrence of multiple Andreev reflection processes in pinholes present in the MgO barrier. 
This mechanism causes the current to flow in large charge quanta ({\em Andreev clusters}), with a voltage 
dependent average value of $ 
m \approx  1+ \frac{2 \Delta}{eV} $ times the electron charge. Because of this charge enhancement effect, the 
shot noise is increased by the factor $m$. 
\end{abstract}

\vskip 0.2cm 
PACS numbers: 72.70.+m, 74.50.+r, 74.80.Fp 
\vskip 0.5cm 
A dc current $I$ flowing through a vacuum tube or a tunnel junction generates shot noise, time dependent 
fluctuations of the current due to the discreteness of charge carriers. Shot noise studies provide information on 
the nature of conduction  not obtainable by conductance studies, e.g. the electric charge of carriers or the degree 
of correlation in the conducting system. For an uncorrelated system in which the electrons do not interact, the 
passage of carriers can be described by a Poisson distribution.  The spectral density of current fluctuations 
$S_I$ then equals full shot noise: $ S_I = 2 q I = P_{\mbox{\tiny{Poisson}}}$ for zero frequency and 
temperature\cite{RogovinScalapino}. The charge quantum $q$ is normally the electron charge $e$.  
 
In superconductor - normal metal (SN) systems Andreev reflection occurs, causing an effective charge to be 
transferred of $2 e$. Due to this doubling of the charge, the shot noise in such a system is predicted to have a 
maximum of twice the Poisson noise\cite{Khlus,MdeJong,Musik}. More recently, giant shot noise in the 
supercurrent is predicted in a single-channel superconductor - normal metal - superconductor (SNS) point 
contact\cite{Averin} which is attributed to transport of large charge quanta ($q \gg e$) at finite voltages caused 
by Multiple Andreev Reflection (MAR)\cite{BTK}. Observation of enhanced charge quanta in SN or SNS 
structures requires a combination of conductance and shot noise measurements. Despite extensive theoretical 
work, experimental results in this field are rare. A recent experiment\cite{Misaki} is performed on a 
NbN/$c$/Nb structure in which $c$ is assumed to be a Nb constriction with a length of 7~nm and a diameter of 
15~nm. At 9.5~K the structure acts like an NS interface but doubled shot noise is not observed. The predicted 
giant supercurrent shot noise is not observed either (at 4.2~K). 
 
From an applied point of view, shot noise in superconducting structures is of interest since this noise forms a 
major limitation to the sensitivity of NbN/MgO/NbN Superconductor-Insulator-Superconductor (SIS) THz 
radiation detectors\cite{DielemanTHz}.  
 
For these reasons, we have investigated quasiparticle current transport and shot noise in an SNS structure in 
which the quasiparticle current is carried by MAR. Anticipating the experimental shot noise results presented 
in this paper we demonstrate that {\em in a system in which multiple Andreev reflections occurs the 
quasiparticle shot noise at $V < 2 \Delta /e$ has a maximum value given by $ S_I =  (1+ \frac{2 \Delta}{e V}) 
2 e I$ because current is effectively carried by multiply charged quanta.} The maximum shot noise is obtained 
if the transmission probability of the system approaches zero\cite{MdeJong}. 
 
\begin{figure}[bt] 
\centerline{\psfig{figure=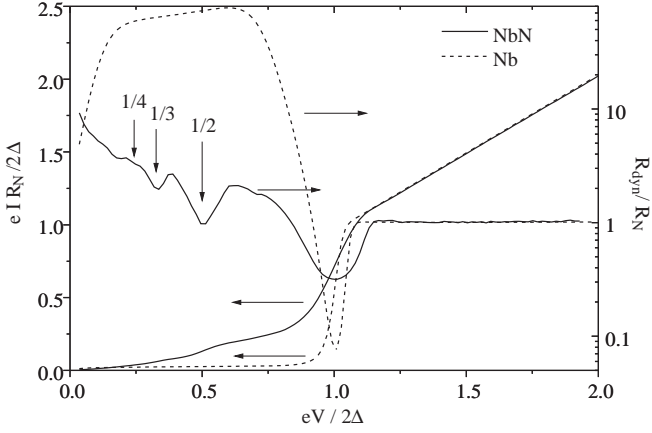,width=86mm}} 
\caption{\label{IVNbNbN} Normalized $I,V$ and $R_{dyn} = dV/dI$ curves of a Nb and NbN junction. The 
subgap structure is clearly visible up to the fourth harmonic in the NbN $R_{dyn}$ curve.} 
\end{figure}

The~structure~under~study~consists of~a NbN/MgO/NbN SIS tunnel junction with small defects in the 
1~nm thick MgO barrier acting as parallel SNS point contacts. Commonly used Nb/AlO$_x$/Nb tunnel 
junctions are known to exhibit very low subgap "leakage" current because the barrier is formed by thermal 
oxidation of a thin Al layer which wets the Nb surface. For NbN junctions such a thermal oxidation process is not 
yet established. Currently the best results are obtained by direct deposition of the barrier material\cite{Shoji}, 
and defects occur 
due to "missing atoms" since the barrier is comprised of merely 1 -- 2 atomic layers in high current density 
junctions\cite{Kleinsasser}.
 In Fig.~\ref{IVNbNbN} the nearly perfect 
tunneling $I,V$ characteristic of a Nb junction is compared with that of a typical NbN junction. Both are 
measured at 4.2~K. The NbN junction area is 0.8~$\mu$m$^2$, the normal resistance $R_N$ is 
25~$\Omega$, the gap voltage $2\Delta/e$ is 4.8~mV. The Nb junction has an area of 
1~$\mu$m$^2$, a normal resistance $R_N$  of 56~$\Omega$ and a gap voltage of 2.8~mV.
Clearly the subgap current of the NbN junction is much larger than the current in the Nb 
junction. In addition the subharmonic gap structure in the NbN $I,V$ and $dV/dI$ curves indicates that the 
subgap current is predominantly carried by MAR. \cite{BTK}. Since MAR is evidence for the presence of 
higher order processes\cite{vanderPost} the observation of MAR indicates conducting paths with 
transmissivities close to 1, which we further call pinholes.
A remarkable feature of a pinhole system is that varying the barrier thickness changes only the number of 
pinholes without affecting the pinhole conductance\cite{Kleinsasser}. Therefore the pinholes can be considered 
as reproducible barrier defects with identical transmission probabilities. From the relative height of the current 
steps at the subharmonic gap structure in Fig.~\ref{IVNbNbN} a pinhole transmission probability $T$ of 0.17 
is derived\cite{vanderPost}. The supercurrent (Fraunhofer)  dependence on magnetic field is nearly ideal, 
indicating a large number of pinholes with a homogeneous distribution over the barrier. All measured
high current density NbN/MgO/NbN junctions exhibit similar behavior.

\begin{figure}[bt] 
\centerline{\psfig{figure=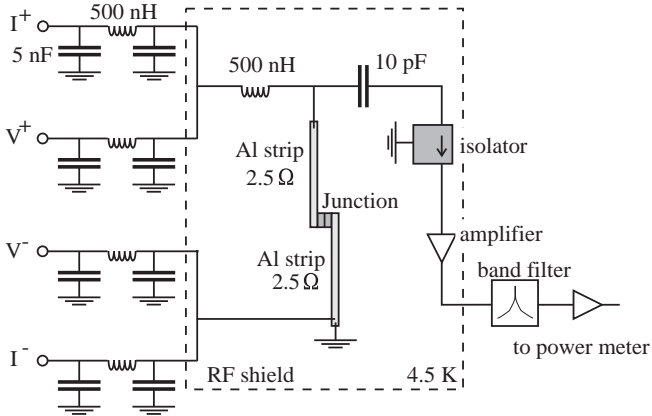,width=86mm}} 
\caption{\label{measuresystem} Schematic layout of the current-voltage and noise measurement system. All 
leads going into the dewar are low-pass filtered  except for the 1.5~GHz wiring. The Al leads suffice as heat 
sink.} 
\end{figure}

 
The noise measurement setup is schematically depicted in Fig.~\ref{measuresystem}. 
All DC leads are filtered, 30~$\Omega$ manganin wires carry currents between room temperature parts and those at cryogenic 
temperatures. The isolator is used to dissipate power reflected off the amplifier into a 50~$\Omega$ load in 
order to avoid gain variations with varying junction output impedance. A band pass filter transmits an 85~MHz 
band centered at 1.5~GHz. Although a spectrum of the signal was not taken, we assume the measured noise to be white 
since 1.5~GHz is high enough to ignore $1/f$ noise of the device. 
  
\begin{figure}[bt] 
\centerline{\psfig{figure=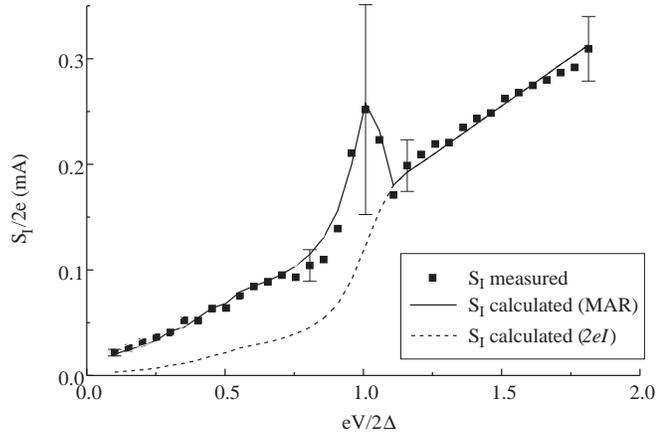,width=86mm}} 
\caption{\label{SINbN} Measured and calculated shot noise generated by the NbN junction, all normalized to 
current by division by $2 e$. The square dots give the measured noise, the dashed line is the calculated Poisson 
shot noise curve. The calculation of the $S_I$ curve labeled (MAR) is explained in the text. Above the gap 
voltage the shot noise follows Poisson noise since current transport through the barrier dominates.  
The error bars, given at specific points to avoid crowdedness of the figure, reflect the uncertainty 
in the noise contributions of the measurement equipment. Errors at intermediate voltage values can be 
roughly linearly interpolated. Offsets in current and noise are negligible since at negative voltage bias 
the same results are obtained.} 

\end{figure}

The total gain of the amplification section is measured by mounting the high quality Nb SIS junction shown in 
Fig.~\ref{IVNbNbN} in the 
measurement setup. Since the Nb junction exhibits single particle tunneling, the shot noise generated by the 
current in the junction can be accurately calculated from the $I,V$ curve\cite{DubashMTT}. Corrections due to 
the measurement frequency of 1.5~GHz are negligible for voltages $V \gg \hbar \omega \approx 0.05 $mV. 
The effect of finite temperature below 0.5~mV is taken into account by using $S_I =  2 e I\coth(e V/2 k_B T)$ 
to calculate the noise  coming from the junction. Comparing this noise with the noise power after filtering 
and amplification gives the 
total gain as well as the noise added by the isolator, cables and amplifier. The output power is given by\cite{DubashMTT} 
\begin{eqnarray} 
\label{EqNbnoisecal} 
\! \! \! \! \! \! \! \!P_{out} & ~=~&  G_{amp}  B ~ (\frac{1}{4} S_I R_{dyn} (1-\Gamma^2) G_{iso}   ~+~    k_B T \, 
\Gamma^2   \nonumber \\ 
& ~+~ &  k_B (1 - G_{iso})  T ) + P_{amp}   
\end{eqnarray} 
in which $G_{amp}$ is the amplifier gain, $B$ is the bandwidth, $I$, $V$ and $R_{dyn}$ are the current, 
voltage and the differential resistance $dV/dI$ of the junction, $\Gamma$ is the reflection coefficient 
$|R_{dyn}-R_{amp}|/(R_{dyn}+R_{amp})$, $G_{iso}$ is the isolator gain, and $P_{amp}$ stands for the 
output noise of the amplifier chain. From the measured and calculated noise power curves the gain and noise of 
the amplifier and isolator can be accurately determined\cite{Netty} using Eq.~\ref{EqNbnoisecal}. The loss of 
the isolator plus cables is 0.25~dB. The total amplification including cable losses and reflections is 80~dB.  
\begin{figure}[bt] 
\centerline{\psfig{figure=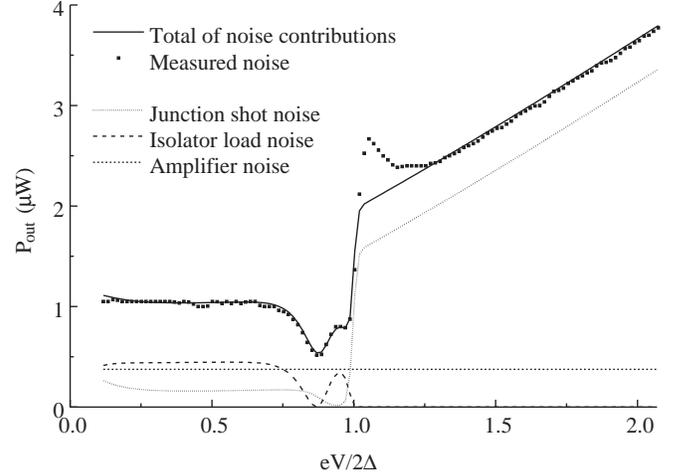,width=86mm}} 
\caption{\label{Nbnoise} Contributions of the Nb junction, isolator, and amplifiers to the total output noise. 
The measured total noise power is accurately given by Eq.~1. The deviation above the gap voltage may be 
connected to a proximity effect layer near the tunnel barrier causing doubled shot noise.} 
\end{figure} 
\newpage 
The shot noise $S_I$ generated in the NbN junction is obtained by performing a measurement of $P_{out}$, 
$I$ and $dI/dV$ as function of voltage similar to that conducted for the Nb junction. Since the loss and gain of 
each component is known, the shot noise $S_I$ can be derived from the output noise $P_{out}$ using 
Eq.~\ref{EqNbnoisecal}. The result is plotted in Fig.~\ref{SINbN} together with the Poisson shot noise $2e I$ 
calculated using the measured $I,V$ curve. The shot noise is expressed as a current by dividing the measured 
$S_I$ by $2 e$ to allow easy comparison with the shot noise calculated form the $I,V$ curve. If the junction 
behaves as an ideal shot noise source as the Nb junction does, a comparison of $S_I / 2 e $ with the dc current 
$I$ will find them identical at voltages above 0.5~mV. Clearly, below the gap voltage the measured noise is much larger than Poisson 
noise.  
We explain this excess noise  by taking into account multiple Andreev reflection processes in the pinholes. For 
an NS structure with a disordered region of length $L$ which is much smaller than the mean free path $l$ in 
this region, the shot noise at $V \ll \Delta / e$ is given by:\cite{MdeJong} 
\begin{equation} 
\label{NSNoise} 
P_{NS}(T) ~=~ \frac{8 (1 - T)}{(2 - T )^2} \cdot P_{\mbox{\tiny{Poisson}}} 
\end{equation} 
in which $T$ is the transparency of the barrier between the normal- and superconducting region. For $T \ll 1$ 
$P_{NS} = 2 P_{\mbox{\tiny{Poisson}}} $ indicating uncorrelated current of particles with charge 2e due to 
Andreev reflection. Realizing that via multiple Andreev reflection much larger charge quanta can be 
transferred\cite{Averin}, the shot noise generated in pinholes by MAR is derived in the following way. The 
pinhole is assumed to consist of two SN and NS structures in series. If the measured transmission 
probability of 0.17 is inserted into Eq.~\ref{NSNoise} the shot noise suppression is a mere 1~\%, below the 
measurement accuracy. Therefore if the effectively transferred  charge $q$ is known, the shot noise is simply 
found from $S_I = 2 q I$. We calculate the charge $q(V)$ in the limit of unity pinhole 
transmission probability using the trajectory method employed in the original paper on MAR\cite{BTK}. This 
enables separate calculation of the currents $I_m$ carried by $m$ Andreev 
reflections. The $I_m$ values form weight factors used to calculate the average transferred charge carried 
by what might be called {\em Andreev clusters}. 
 
\begin{figure}[bt] 
\centerline{\psfig{figure=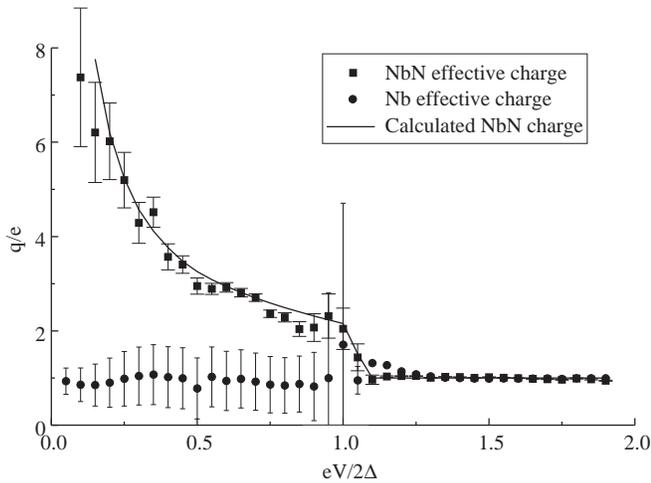,width=86mm}} 
\caption{\label{effcharge}Effective charge as function of voltage. The square dots give the charge derived 
from the measurements. The calculated charge values are used to calculate the shot noise in Fig.~\ref{SINbN}.
 For comparison the effective 
charge of the Nb calibration junction is shown by circles.} 
\end{figure}

The total current flowing from the left superconductor (at voltage V) to the right superconductor (at zero 
voltage) is given by the difference of\cite{BTK} 
\begin{eqnarray} 
\lefteqn{\! \! \! \! \! \! \! \! I_{\mbox{\tiny LR}}  = ~  \frac{1}{e R_0} ~ \int_{- \infty}^{\infty} dE f_0(E - eV) 
[1 - A(E - eV)]   \cdot}  \nonumber  
\end{eqnarray} 
\begin{eqnarray}\label{BTKLRcurrents} 
\lefteqn{\! \! \! \! \! \! \! \! [1 + A(E) + A(E)A(E + eV) + \cdots]} 
\end{eqnarray} 
and 
\begin{eqnarray} 
\lefteqn{\! \! \! \! \! \! \! \! I_{\mbox{\tiny RL}} ~=~ \frac{1}{e R_0} ~ \int_{- \infty}^{\infty} dE f_0(E) [1 - 
A(E)] \cdot  }\nonumber  
\end{eqnarray} 
\begin{eqnarray} 
\label{BTKRLcurrents} 
\lefteqn{\! \! \! \! \! \! \! \! [1 + A(E - eV) + A(E - eV)A(E - 2eV) + \cdots]} 
\end{eqnarray} 
where $R_0$ is the total normal state resistance of the pinholes, $f_0(E)$ is the Fermi distribution and $A(E)$ 
is the energy dependent Andreev reflection probability. The first term $[1- A]$ denotes the fraction of available 
electrons in the superconductor which is transmitted into the normal region. The last term $[1 + \cdots]$ gives 
the transferred charge multiple which is 1e if no Andreev reflections take place, 2e for one Andreev reflection, 
and so on. The currents $I_m$ ($m = 1, 2, \ldots$) carried by $m$-electron processes are calculated by splitting 
Eqs.~\ref{BTKLRcurrents} and \ref{BTKRLcurrents} into the $m$-electron parts. For example $I_2$ is given 
by 
\begin{eqnarray} 
\lefteqn{\! \! \! \! \! \! \! \!  I_2(V)  =  \frac{2}{e R_0}  \int_{- \infty}^{\infty} dE \mbox{\Large (} ~f_0(E - eV) 
[1 - A(E - eV)]  \cdot   ~~}\nonumber  
\end{eqnarray} 
\begin{eqnarray} 
\lefteqn{\! \! \! \! \! \! \! \! A(E) [1 - A(E + eV)]  - } \nonumber   
\end{eqnarray} 
\begin{eqnarray}\label{BTKIn} 
\lefteqn{\! \! \! \! \! \! \! \! f_0(E) [1 - A(E)]  A(E - eV) [1 - A(E - 2eV)] ~\mbox{\Large )}} 
\end{eqnarray} 
Since the current $I_m$ is carried only by carriers with an $m$-fold charge, the magnitude of this current 
divided by the total current gives the relative contribution to the average transferred charge. Therefore the 
average charge~$q$ of an {\em Andreev cluster} for a given voltage $V$ is given by:  
\begin{equation} 
\label{Averagechargesummation} 
q(V)~~=~~   \frac{ \mbox{$ \sum_{m=1}^{\infty}  m \cdot I_m(V) $}}{\mbox{ $\sum_{m=1}^{\infty}I_m(V) 
$}} \cdot e  
\end{equation} 
The resulting charge-voltage curve is shown in Fig.~\ref{effcharge} and is used to calculate the noise $S_I = 
2 q(V) I$ in Fig.~\ref{SINbN}. The measured charge values in Fig.~\ref{effcharge} are 
obtained by dividing the measured shotnoise by $ 2 e I$. The similarly derived unity charge of the Nb junction is 
shown for comparison. The effect of finite temperatures on the shot noise correction factor $\coth({eV}/{2k_B T})$  at 
voltages $V \ll 2\Delta/e$ is negligible since $\coth{({q(V) V}/{2 k_B 
T})} \approx \coth{((1+ \frac{2 \Delta}{e V}) \cdot \frac{eV}{2 k_B T})} \approx \coth{(\frac{\Delta}{k_B 
T})} \approx 1$. 
    
Apart from a superposed subharmonic gap structure 
the effective NbN charge is proportional to $1/V$. This can be appreciated by performing  the 
charge calculation ignoring Andreev reflection for energies $|E| > \Delta$. Higher order terms 
vanish and the average charge is equal to the charge quantum transferred; $q_n =  n e$ at voltages 
$V_n = \frac{2 \Delta}{n -1}$ for $n = 2, 3, \dots$ giving an analytical approximation $q(V) = e + \frac{2 
\Delta}{V}$. 

In conclusion, for the first time shot noise much larger than Poisson noise is observed in  an SIS structure. 
A NbN SIS tunnel junction with pinholes is employed to obtain current transport dominated by multiple 
Andreev reflections (MAR). Due to the occurrence of MAR large charge quanta are transferred between the 
electrodes, causing a significantly enhanced shot noise. A simple model is developed to calculate the effective 
charge from which the shot noise is obtained. The model answers the question about the origin of 
excess noise in NbN\cite{DielemanTHz} and very high current density Nb\cite{GertAPL} SIS junctions in THz 
applications. 
 
{\em Note added in proof.} After submission of this manuscript we became aware of theoretical 
work\cite{Hessling} which analyzes MAR enhanced shot noise, in particular the dependence of $S_I$ on 
temperature and transmission. In our semi-empirical theory we circumvent the problem of calculating $S_I$ by 
calculating $q(V)$ (ignoring coherence effects) and using the measured $I$.

 
We thank N.~Whyborn for a remark initiating this work and N.B.~Dubash for sharing his data and ideas. 
Helpful discussions and general assistance of  S.G.~den~Hartog, B.J.~van~Wees, J.B.M.~Jegers, J.R.~Gao, 
H.~Golstein, H.~van~de~Stadt, W.~Hulshoff, D.~Nguyen, and H.H.A.~Schaeffer are acknowledged. This work 
is supported in part by the European Space Agency under contract No. 7898/88/NL/PB(SC) and the Stichting 
voor Technische Wetenschappen.

\end{document}